\documentclass[prd,nofootinbib,preprint]{revtex4}
\usepackage{graphicx, epsfig}
\usepackage{color}
\usepackage{mathrsfs}
\newcommand{\scri}{\mathscr{I}}
\usepackage{graphics}
\usepackage{bm}

\begin{document}

\title{
On Constructing Baby Universes and Black Holes
}

\author{
Tanmay Vachaspati
}
\affiliation{
CERCA, Department of Physics, Case Western Reserve University,
10900 Euclid Avenue, Cleveland, OH 44106-7079, USA.\\
tanmay@case.edu}

\begin{abstract}
\noindent
The creation of spacetimes with horizons is discussed, focussing on
baby universes and black holes as examples. There is a complex interplay
of quantum theory and General Relativity in both cases, leading to
consequences for the future of the universe and the information loss 
paradox, and to a deeper understanding of quantum gravity.
\end{abstract}

\maketitle

General Relativity allows for a spacetime to have horizons, that is,
spacetime regions that are inaccessible to certain observers. 
Due to quantum effects, horizons have novel consequences
such as Hawking radiation from black holes \cite{Hawking:1974sw}, 
and ``freezing'' of quantum fluctuations in the inflationary 
vacuum \cite{Mukhanov:1981xt}. These
quantum effects are of fundamental importance and interest. 
Quantum fluctuations in inflationary cosmology may be responsible 
for seeding galaxies and cosmic large-scale structure, while
Hawking radiation leads to the evaporation of black holes and
to the information loss paradox.
In this essay, we will discuss how to create two systems that have 
horizons, namely baby universes and black holes, the difficulties 
that are encountered in this process, and their possible resolution. 

First consider the tantalizing possibility of creating a 
baby universe in a laboratory \cite{Farhi:1986ty,Fischler:1989se,
Farhi:1989yr,Fischler:1990pk,Borde:1998wa}. At the classical level 
all that we need do is to produce a large enough bubble of false 
vacuum, after which it will inflate and become a baby universe all 
on its own. This appears to be simple, but to successfully complete 
the process without the production of singularities, it can be shown 
that the process needs a matter source that 
violates the null energy condition. Classical matter sources
that violate the null energy condition are not known to exist and, 
on the contrary, there are arguments to show that the existence of 
such sources would lead to unphysical consequences such as closed 
timelike curves. So a baby universe cannot be produced in the 
laboratory with (known) classical matter sources. This conclusion 
need not disappoint us, however, since the real world 
is quantum and even rather ordinary quantum fields can be shown to 
violate the null energy condition with no untoward consequences 
\cite{GutVacWin??,Winitzki:2001fc,Vachaspati:2003de}. 
Perhaps quantum physics can be utilized to produce baby universes? 

\begin{figure}
\scalebox{0.50}{\includegraphics{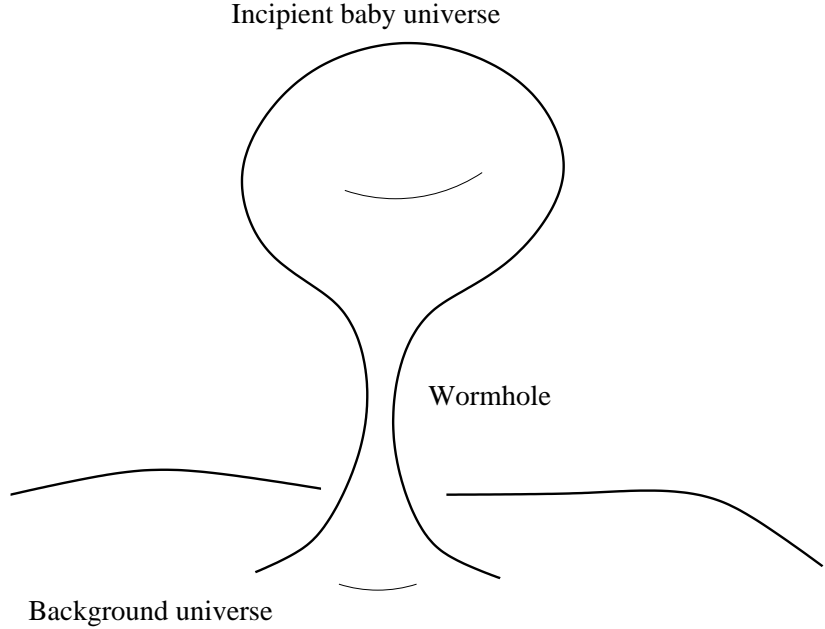}}
\caption{\label{babyu}
Embedding diagram for an incipient baby universe connected 
to the background universe by a wormhole that needs null energy 
condition violating
sources to keep it from collapsing and pinching.
}
\end{figure}

Imagine trying to produce a baby universe in a laboratory via the
false vacuum bubble. The bubble will be connected to the laboratory 
by a wormhole, somewhat like an umblical cord as in 
Fig.~\ref{babyu}. As long as there are null energy condition violating 
quantum fields present in the wormhole, the wormhole stays 
open. However, once the null energy condition violation ceases, 
the wormhole collapses, 
leading to a singularity, and the future evolution of the spacetime
cannot be predicted. Since quantum null energy condition violations are 
short-lived, the existence, let alone the properties, of the baby 
universe cannot be calculated. Thus baby universes, with their
horizons, cannot be shown to form in the laboratory, even when 
quantum effects are taken into account.

Baby universe production is closely tied to the initial conditions 
necessary for inflation. In inflationary cosmology, a region of
spacetime undergoes super-luminal expansion within a background
cosmology. If the inflating region is initially small compared to 
the background horizon, the process is very similar to the creation
of a baby universe in a laboratory since effectively a bubble of
false vacuum needs to be created that can then start inflating. 
Since we have seen that baby universes cannot be produced in the 
laboratory, it also implies that inflationary regions cannot be 
produced on sub-horizon scales, such as our laboratory. This implies 
that the false vacuum dominated bubble that grows by inflation and 
produces all the structure that we see, must start out larger than 
the initial horizon size \cite{Vachaspati:1998dy}. The fact that 
the initial bubble has to be super-horizon size means that inflation 
must be preceded by a causation that extends beyond the light cone.

The argument showing that baby universes cannot be produced in
the laboratory changes if quantum effects occur on a cosmological 
scale and the wormhole is as big as the background horizon. Then 
there is a pre-existing cosmological horizon and the expansion of 
the background universe can prevent the wormhole from collapsing 
and the concomitant singularities from forming. This is the basis 
of several cosmological models
with eternal regeneration of new universes, such as in the 
eco-friendly ``recycling universes'' \cite{Garriga:1997ef} and 
the minimalistic ``island cosmology'' \cite{islands}. These
cosmological models reveal a new paradigm where cosmic habitats
and humanity are eternal, and provide an escape from the dreary
conclusion that our universe has a bleak and empty future because
of the accelerating expansion rate. 

Horizons are also discussed in the context of gravitational collapse. 
At first sight, it appears that gravitational collapse inevitably 
leads to black hole formation which is accompanied by formation of 
an event horizon. However, here too, quantum effects play a subtle 
role during the collapse process, that may prevent the formation of 
an event horizon and possibly provide a resolution of the information 
loss paradox. 

Before looking at the gravitational collapse problem, consider
a more mundane problem where a spacecraft needs to deliver 
fuel to a distant destination like Mars. Suppose that the spacecraft
only has a single tank for fuel, to be used for its own journey as 
well as to deliver to its destination. Then, as the spacecraft flies,
it consumes the very fuel that is to be delivered. Further, the
spacecraft can successfully deliver fuel to its destination, only
if it burns less fuel on its journey than that initially contained 
in its tank. 

Gravitational collapse is similar to the spacecraft's journey.
Suppose there is a spherical shell of matter that is collapsing
toward forming a black hole. Just as the spacecraft is on a
journey to deliver fuel to Mars, the collapsing shell is on a journey 
to deliver mass to a central region that is compact enough to form 
a black hole. Again, in analogy with the spacecraft, the shell burns
its mass and steadily loses energy due to quantum effects as it 
collapses \cite{VacStoKra,Boulware:1975fe,Gerlach:1976ji,Hajicek:1986hn,
Alberghi:2001cm, Ashtekar:2005cj}.
(The radiation is very similar to Hawking 
radiation but it does not require a black hole to be present.) In a 
slight departure from the analogy, as the shell evaporates, its mass 
gets smaller, and its ``destination'' moves further away since it 
needs to become even more compact to form a black hole. The shell 
keeps chasing its destination but possibly never gets there, as 
in a mirage \cite{Gerlach:1976ji,VacStoKra}.


The General Relativistic problem is, however, a little more
subtle than the spacecraft problem because there is freedom to 
choose coordinates, and in particular, the time slicing. 
Instead it is unambiguous to think in 
terms of observation of two different events, one signalling the 
evaporation of the collapsing wall and the other the formation 
of a black hole. The event signifying evaporation may be taken 
to be when an external observer, who is monitoring the emitted
radiation, finds that all the initial energy of the shell has
been burned up. The formation of a black hole may be signalled
by the disappearance of some object. Then the question is if
objects are observed to disappear before the total energy is 
burned up. 

We know that the gravitational redshift of light emerging from 
a flashlight just outside a Schwarzschild event horizon is very 
large, and diverges in the limit that the flashlight approaches 
the horizon. Therefore, if the metric outside the
shell, which is an incipient black hole, has the Schwarzschild form,
then an object will never be seen to fall through the black hole
event horizon by an external observer. Yet the external
observer will collect the total mass of the gravitationally
collapsing shell in a finite time, indicating that the
shell burns up before a black hole is formed. 

\begin{figure}
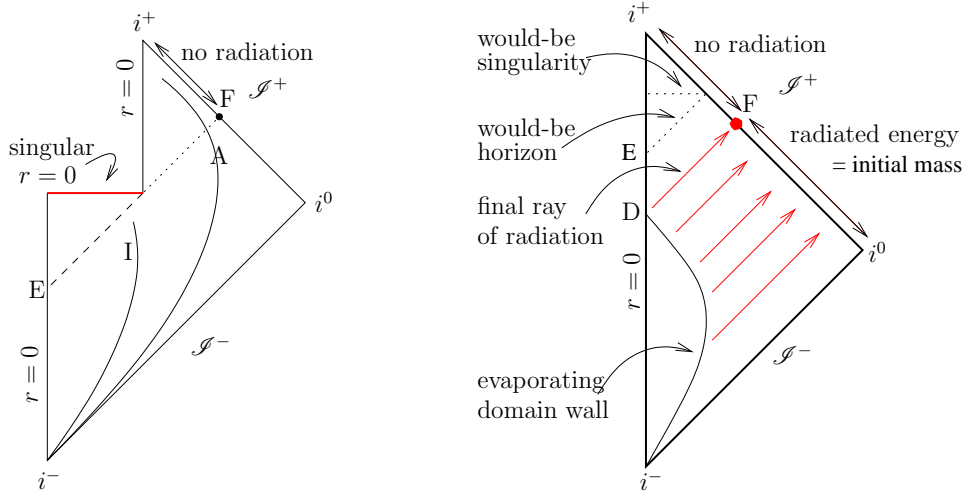

\centerline{\scalebox{0.80}{\input{usualbh2.pstex_t} 
       \hskip 1 in                     \input{minkowski2.pstex_t}
}}
\caption{
A collapsing shell can either succesfully become a black hole
leading to the spacetime picture on the left, or else it can
burn up by pre-Hawking radiation and lead to the picture on
the right. The picture on the left leads to the information
paradox, while the picture on the right implies that no
black holes are formed. It is also possible that the correct
spacetime picture depends on the mass of the collapsing object.
}
\label{minkowski}
\end{figure}

We have used the Schwarzschild metric to argue that an object falling 
into a gravitationally collapsing object is never seen to disappear. 
Using Birkhoff's theorem, the Schwarzschild form is inescapable.
However, the Schwarzschild metric is derived from the classical 
Einstein equations, without taking quantum effects and radiation 
backreaction on the metric into account. In the 
case of the spacecraft bound for Mars, if we want to calculate the 
total fuel that will be burned, we should take into account the fact 
that the spacecraft gets lighter as the fuel is burned, and hence 
requires less fuel to accelerate. Similarly, as the collapsing shell 
loses mass, its collapse and the spacetime around it get affected.
Although it is hard to see how the metric of the incipient black
hole could be anything other than of Schwarzschild form, until
the problem is solved including backreaction, we cannot really
be sure that the black hole event horizon does not form, and we are 
left with the two possible spacetime pictures shown in Fig.~\ref{minkowski}.
The first of these pictures is the conventional picture where burn out
does not occur and a black hole is formed. An exciting aspect of the 
second picture where the shell burns out is that it resolves the 
information loss paradox in a very natural way \cite{VacStoKra}.


It is quite remarkable how fundamental a role quantum physics plays
in constructing systems with horizons. Any attempt at making baby 
universes and black holes, must face the constraints imposed by 
quantum effects and also utilize the opportunities offered
by these very effects. To understand the ``yin and yang'' of
quantum physics in spacetimes with horizons is to get that much 
closer to an understanding of quantum gravity.

\begin{acknowledgments}
This work was supported by the DOE and NASA.
\end{acknowledgments}

\end{document}